# Cellular Automata on Quantum Annealing Systems

Robert A. Dunn
Lockheed Martin Corporation, USA

We present herein an introduction to implementing 2-color cellular automata on quantum annealing systems, such as the D-Wave quantum computer. We show that implementing nearest-neighbor cellular automata is possible. We present an implementation of Wolfram's cellular automata Rule 110, which has previously been shown to be a universal Turing machine, as a QUBO suitable for use on quantum annealing systems. We demonstrate back-propagation of cellular automata rule sets to determine initial cell states for a desired later system state. We show 2-D 2-color cellular automata, such as Conway's Game of Life, can be expressed for quantum annealing systems.

**Keywords**: quantum annealing, cellular automata, Rule 110, Game of Life, Turing machines

## Introduction

Cellular automata (CA) systems, which describe the sequential transition between the states of a set of cells subject to a fixed (usually simple) set of rules, may not seem a natural fit for adiabatic quantum computers (AQC) and quantum annealing systems. A single run of an AQC probabilistically converges to a single qubit pattern rather than a sequence of patterns.

However, when a set of qubits is viewed as a collection of multiple sequential states of the cellular automata, then the rules governing the evolution of the cellular automata form the basis of the AQC Hamiltonian describing how the subsets of qubits corresponding to each generation of the cellular automata are related to their neighbors. Nearest-neighbor cellular automata rules can be particularly amenable to AQC implementation given the limited qubit connectedness on current hardware graphs.

On AQC systems, the ground state of the Hamiltonian represents self-consistent successive generations of cells under the rules of the given cellular automata.

In this paper, we show that the construction of such AQC Hamiltonians is a straight-forward matter for 2-color cellular automata systems. We demonstrate some common cellular automata rules, including the 1-D 2-color cellular automata Rule 110 [WOLFRAM] and the 2-D 2-color cellular automata Conway's Game of Life.

We then consider the use of AQC systems in the reverse evolution of cellular automata systems to create input data which will evolve to a specified final state.

**1-D, 2-Color, Nearest-Neighbor Cellular Automata**

Following the convention of other works, we define a 2-color cell's state to be "on" as color 1 (shown in black) and "off" as color 0 (shown in white). This choice makes it more natural to express the system Hamiltonians in the form of a quadratic unconstrained binary optimization (QUBO) problem rather than the equivalent Ising model, i.e.

$$H = \sum_{i,j} Q_{ij} x_i x_j \qquad\qquad x \in [0, 1]$$

In displays of successive generations of the 1-D CA, the generations progress from top to bottom. CA patterns shown on a rectangular grid indicate a visualization of the CA rule, whereas an interconnected mesh of nodes is used in this paper to represent the corresponding qubit graph, where the inter-node lines reflect non-zero 2nd order terms in the resultant Hamiltonian. (Ancillary qubits in the Hamiltonian are not shown on this type of display, for clarity.) A representative display of a simple checkerboard pattern CA rule in both display formats is given in Figure 1.

| | |
|---|---|
| 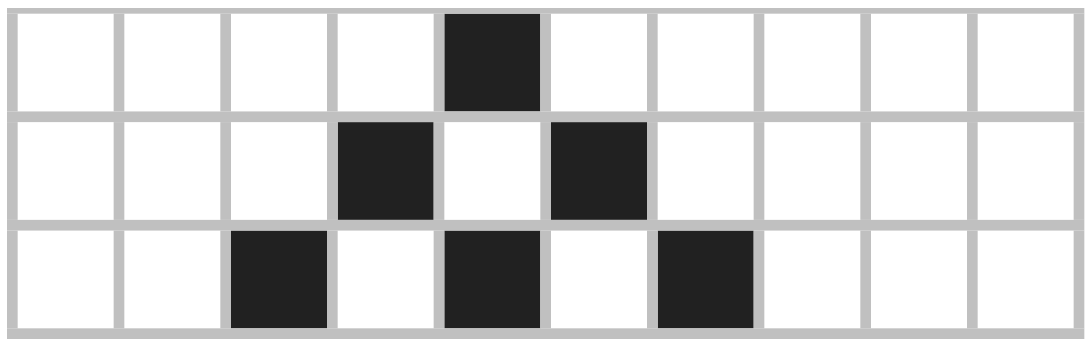 | 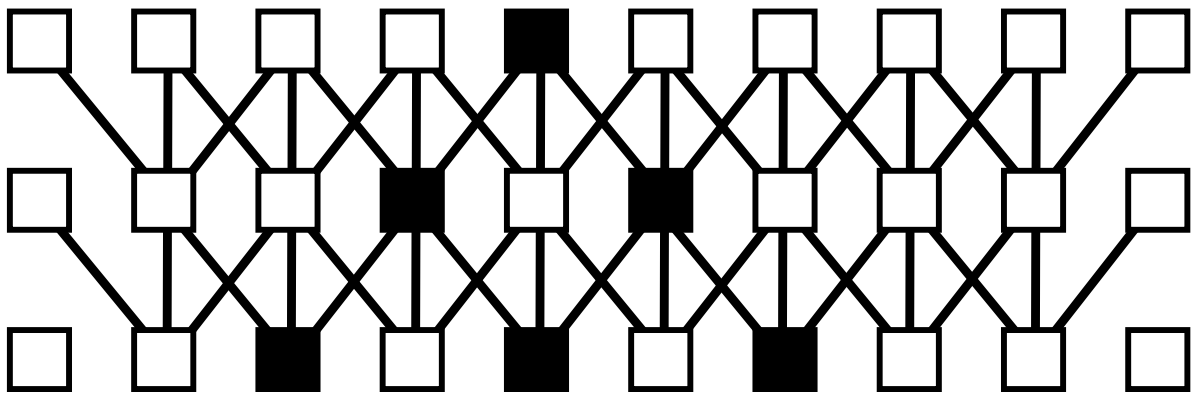 |
| Three generations of a 2-color 10-unit cellular automata evolving under a nearest-neighbor rule. | Thirty qubits arranged to represent three successive generations of a 2-color, 10 unit nearest-neighbor cellular automata. |

*Figure 1: A visualization of a simple "checkerboard pattern" CA rule in both CA and qubit graph formats.*

For any given cell *p*, we denote its state in generation n as $p_n$ (and thus its state in the next generation as $p_{n+1}$). For 1-D CA rules, the cell to the left of *p* is *L* and the cell to the right is *R*. Using this notation, all 1-D nearest-neighbor rules depending only on the current generation's states are then expressed as the value of $p_{n+1}$ as a function of $p_n$, *L*, and *R*, i.e. $p_{n+1} = f(p_n, L, R)$. As an example, the checkerboard pattern shown above can be described as:

$$p_{n+1} = L \vee R$$

Implementing the CA ruleset as a system Hamiltonian is achieved by the piecemeal construction of terms implementing the CA rule as a relationship between the qubits involved in the single-step evolution of a cell -- performed for each cell for each generation.

$$H_{total} = \sum_{g \in generations} \sum_{c \in cells} H_{c,g}$$

The Hamiltonian term for each cell is constructed such that the ground state corresponds to a pattern satisfying the cellular automata rule and has a lower energy than all patterns that do not meet the cellular automata rule. In this manner, the combination of terms for all cells over all generations considered will result in a system Hamiltonian whose global energy minimum corresponds to patterns of cell states that satisfy the cellular automata rule sets. This is most easily achieved when the Hamiltonian term for each cell has a ground state of zero, as is the case for the examples presented herein, such that no weighting terms are required to maintain the integrity of the logical constraints.

**Edge Conditions**

For 1-D cellular automata of finite size (a finite number of cells in each generation), the means of handling the left-most and right-most cells must be considered. These edge cases typically require undefined values from the prior generation. Several conventions are available: considering the missing data to always be 0, always evaluating the edge cell to 0, or applying a cylindrical mapping (rather than the more traditional planar mapping) such that each generation is a closed loop.

For purposes of this paper, we treat 1-D edge cases that require unavailable data to always evaluate to 0. This does not constrain the approach used, and indeed it is trivially shown that other edge conventions are supported in similar fashion.

**The Initial Pattern: A Checkerboard**

The checkerboard pattern is a convenient initial example because its implementation does not involve the use of ancillary qubits to enforce the CA rule set. For each cell, the logic $p_{n+1} = L \vee R$ can be expressed as:

$$H_{cell} = L + R + p_{n+1} + LR - 2Lp_{n+1} - 2Rp_{n+1}$$

Construction of the full system Hamiltonian is formed by the summation of the cell-by-cell Hamiltonian terms, then repeated for each generation. The number of logical qubits to the describe the problem equals the number of total cell states (one per cell per generation) and the Hamiltonian construction requires a polynomial number of steps (as a function of the number of cells and generations studied).

Once the system Hamiltonian has been constructed, the initial conditions may be freely chosen. While it is customary to demonstrate the behavior of the CA ruleset by having only one cell set to 1 in the initial generation, that is neither a requirement of CA systems nor is it mandatory for this AQC Hamiltonian.

The system Hamiltonian approach used in this AQC study easily provides an opportunity that is more bothersome to obtain in traditional digital CA computations, namely that one or more cells in the initial generation may be unset. For the AQC system, this means that the unset cells are in a superposition of 0 and 1 prior to computation and will be determined (if possible) in the valid ground states of the Hamiltonian. We discuss this further later in this paper.

For the checkerboard pattern, each interior cell has interactions with (up to) 6 other cells. This relatively low qubit connectivity is amenable to embedding on current hardware graphs of machines such as the D-Wave 2000Q.

The number of cells and number of generations which can be evaluated in a single run is naturally dependent on the number of physical qubits and the qubit hardware graph available on the AQC system. However, for forward propagation of the CA system, the number of generations studied may be increased by successive runs in which the final generation state of one run is used as the initial generation state of the following run. While that is easily accomplished, we do not suggest that is the most efficient means of studying large numbers of generations of simple CA forward propagation -- a traditional digital computer is likely to be the better solution for most cases.

**Considering Rule 110**

The one-dimensional, two-color, nearest-neighbor cellular automata Rule 110 [WOLFRAM] has been previously shown by [Cook] to be a universal Turing machine. We now consider implementing Rule 110 as an AQC Hamiltonian.

There are many equivalent expressions of Rule 110; we find the following to be a convenient form for the generation rule:

$$p_{n+1} = (\neg (L \wedge p_n \wedge R)) \wedge (p_n \vee R)$$

where L and R are the left and right neighbors, respectively.

It is convenient to utilize three ancillary qubits to store intermediate results in the implementation of this rule:

$$C_1 = p_n \vee R$$

$$C_2 = p_n \wedge R$$

$$D = L \wedge C_2$$

such that the rule becomes:

$$p_{n+1} = \neg D \wedge C_1$$

The evolution rule for cell $p_n \rightarrow p_{n+1}$ can now be expressed in QUBO form via:

$$H_{term} = C_1 + p_{n+1} - C_1 D + 2Dp_{n+1} - 2C_1 p_{n+1}$$

before appending the corresponding ancillary qubit terms. For clarity, we show each of the ancillary qubit expressions on a separate line rather than gathering the polynomial terms for the cell's Hamiltonian:

$$\begin{aligned} H_{cell} = & C_1 + p_{n+1} - C_1 D + 2Dp_{n+1} - 2C_1 p_{n+1} + \\ & p_n + R + C_1 + p_n R - 2p_n C_1 - 2RC_1 + \\ & 3C_2 + p_n R - 2C_2 p_n - 2C_2 R + \\ & 3D + LC_2 - 2LD - 2C_2 D \end{aligned}$$

for which the ground states have H = 0 for the correct evolution of the cell (Appendix B).

The total Hamiltonian shows the expected behavior, as seen on a test grid of 6 generations of 8 cells.

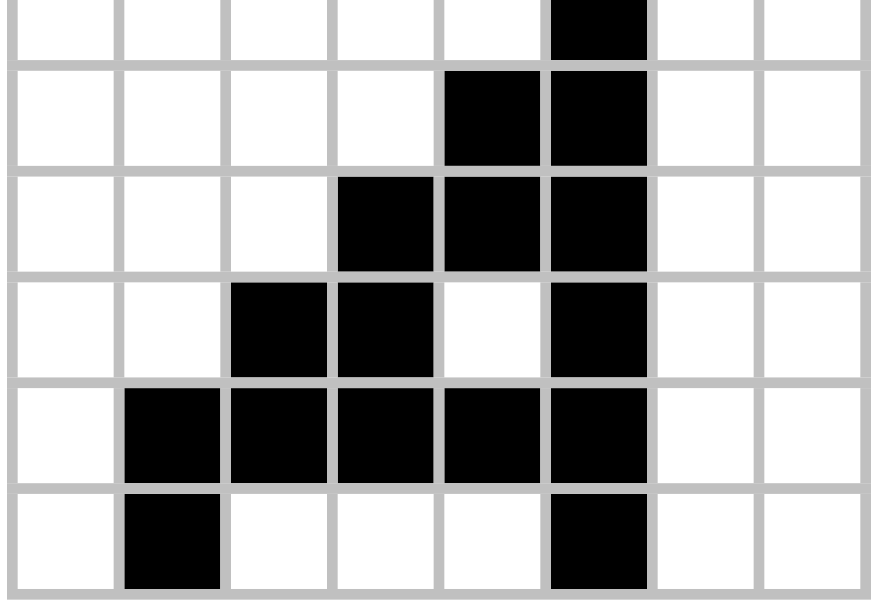

Figure 2: *Six generations of Rule 110, starting from a single 'on' cell.*

**Running on Arbitrary input**

Once a Hamiltonian has been constructed for a given problem, the next consideration is the specification of the input values. With the Hamiltonian in symbolic algebraic form, the input variables may be set by explicit variable substitution and subsequent algebraic simplification. By this technique any variable may be set prior to execution on AQC hardware or simulator.

Because the form of the system Hamiltonian is fixed for a given number of cells and generations under a particular CA rule set, the Hamiltonian may be run for an arbitrary set of input data without further construction. As an example, rather than the single "on" cell of the traditional demonstration, a more complex initial state can be used (Figure 3):

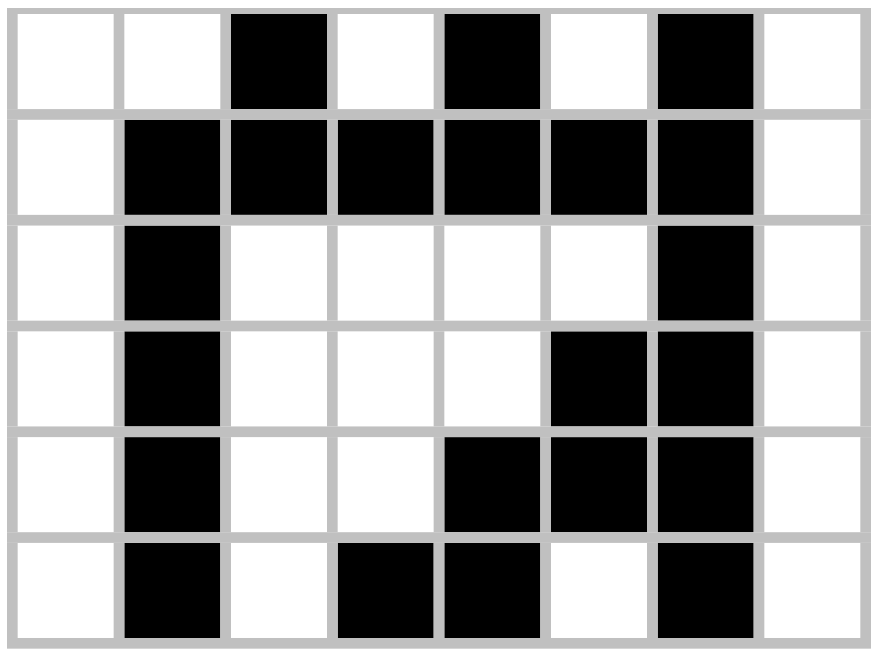

Figure 3: *Running the Hamiltonian for Rule 110 on an initial generation with 3 'on' and 5 'off' cells.*

With the ability to run on arbitrary input data, we have shown that this system Hamiltonian can fully implement the 1-D 2-color cellular automata Rule 110 -- and thereby confirm that the QUBO form suitable for use on an adiabatic quantum computer implements a universal Turing machine.

The QUBO form of this system Hamiltonian may be trivially converted to an Ising model, therefore machines capable of running Ising models on arbitrary input data can also run CA Rule 110 and are thus also confirmed to be universal via this approach.

While this shows that any Turing-computable function could be implemented on an AQC system using cellular automata, we do not suggest this is the most efficient means of programming complex functionality.

The logical qubit graph in the Hamiltonian for Rule 110 may prove problematic for systems with low hardware qubit connectivity. Qubits representing the state of interior cells in middle generations have interactions with 10 other logical qubits, which may be difficult to embed on some systems.

**Forward and Backward**

Because we describe the entire system as a single state, the distinction between 'input' and 'output' is much more flexible than is the case for forward-only propagations of cellular automata rules. Indeed, we may choose any pattern of cell states in any (or all) generations to be the input state of the system and then solve for unknown states which are compatible with both the input state and the rule set in use.

Classically, back-propagation of cellular automata rule sets may be expressed as satisfiability problems. In the quantum annealing approach, the Hamiltonian describing the system does not depend on propagation direction. Accordingly, there is no need to create the satisfiability form or craft a new Hamiltonian -- the existing system Hamiltonian may be used with the final generation provided as the input data.

Let us consider a simple example: a 3-generation case operating under Rule 110 with the $3^{rd}$ generation specified as shown below -- in which only the final generation is known: (Grayed cells represent unknown cell states.)

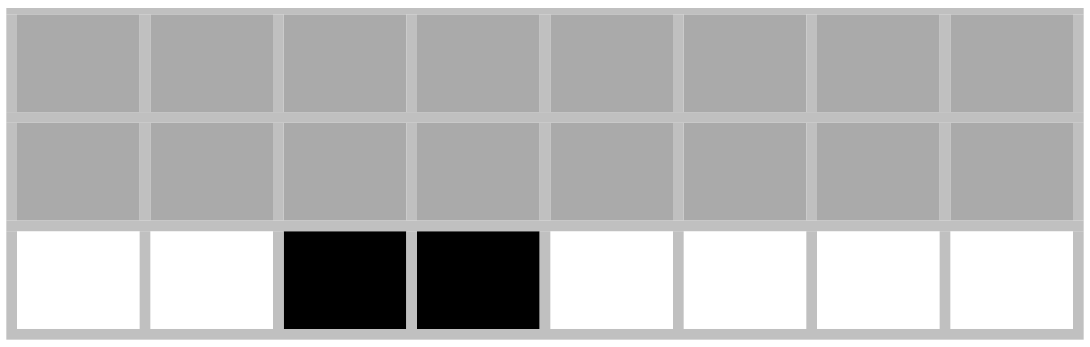

Figure 4: *Three generations of a cellular automata in which the final generation is specified as the input data.*

Setting the value of the cells in the $3^{rd}$ generation, we then run the resultant Hamiltonian. This has a single solution, with H = 0.

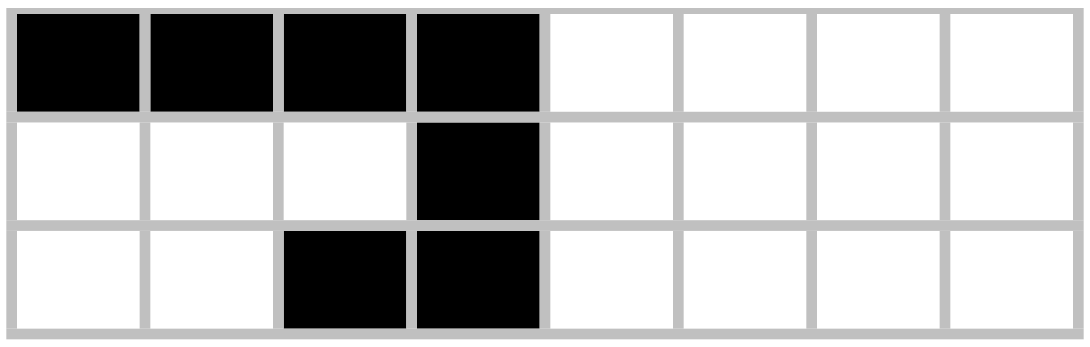

Figure 5: *The unique solution to the cellular automata problem specified in Figure 4.*

As a complementary example, we consider the following $3^{rd}$ generation state as input, again propagating using Rule 110:

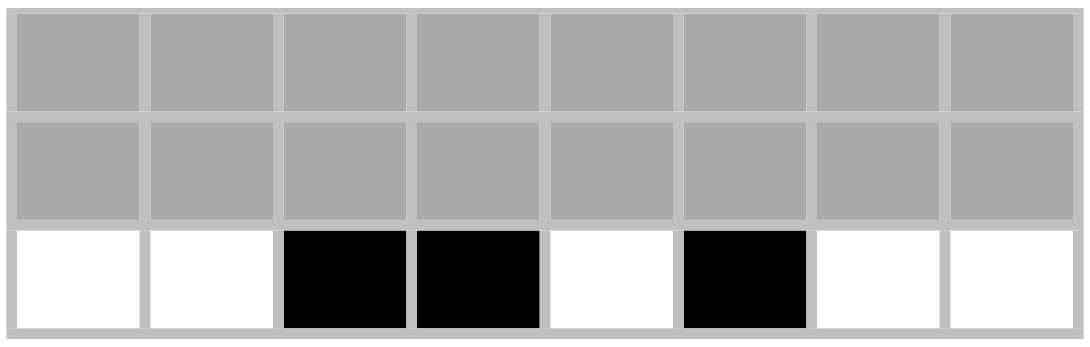

Figure 6: *Three generations of a cellular automata in which the final generation is specified as the input data, for which there are multiple solutions.*

This does not uniquely specify the system under our choice of edge rules, resulting in a degenerate ground state of 4 solutions.

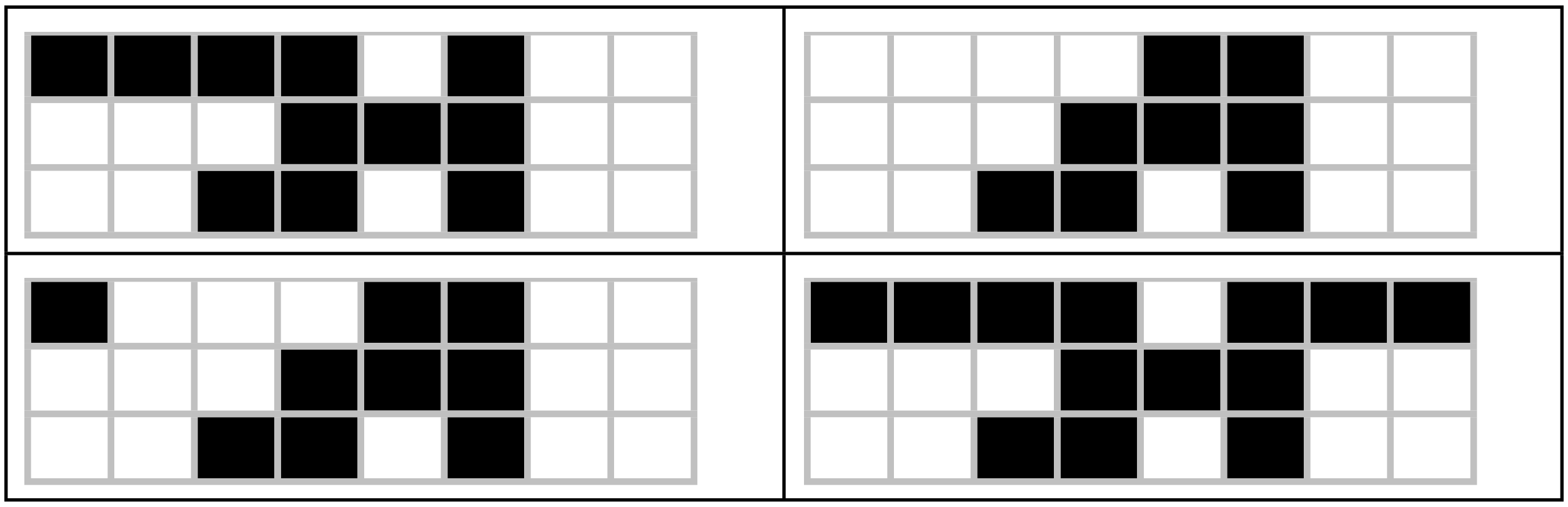

Figure 7: *The degenerate solutions to the cellular problem in Figure 6. There are four input states that evolve to the specified third generation.*

**In Cases of No Solutions**

While forward propagation of the cellular automaton is both possible and deterministic for a well-behaved rule set, backward propagation is not guaranteed to be so. For some cellular automata rules there are output states which cannot be reached from any input state in the given number of generations. It is worth considering the behavior of quantum annealing systems in these scenarios.

Consider a three-generation scenario operating under Rule 110, where the 3$^{rd}$ generation is specified and we seek to back-propagate.

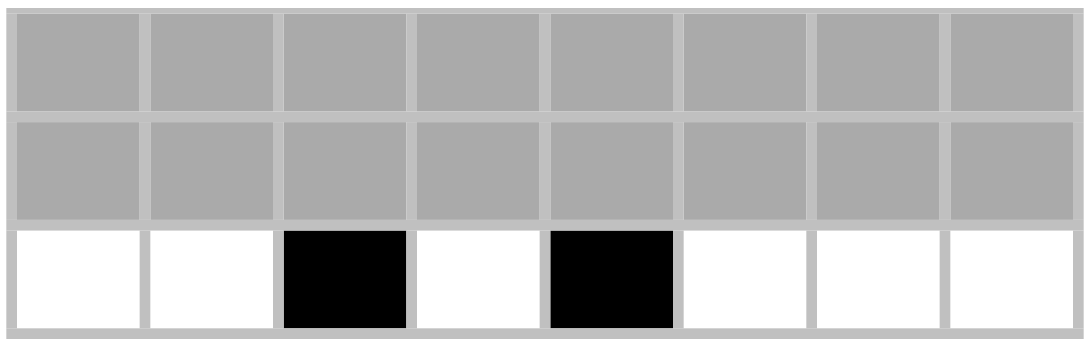

Figure 8: *A three generation cellular automata Rule 110, in which the third generation is specified as input.*

There is no solution to this problem.  When the problem Hamiltonian is solved, there are degenerate solutions:

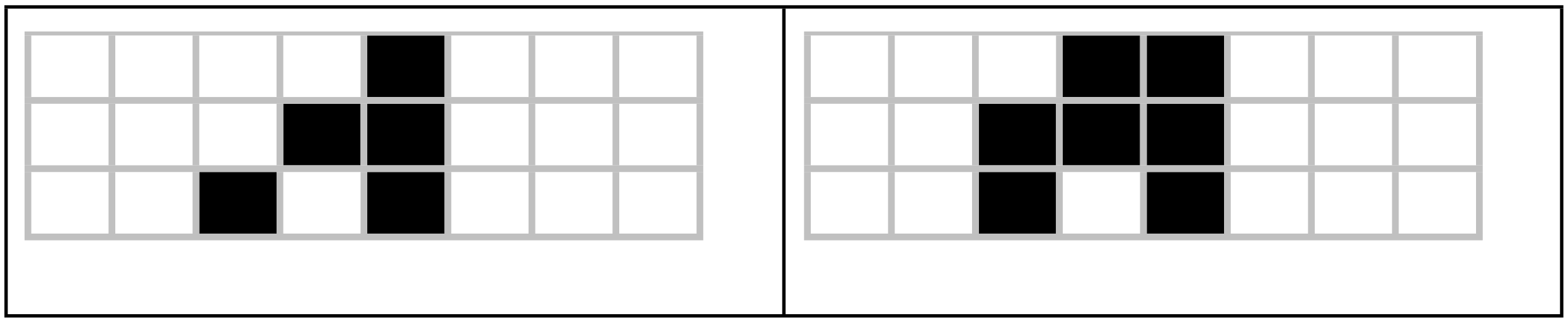

Figure 9: *Two of the lowest energy states of the Hamiltonian for the problem in Figure 8.  No solution is valid under the CA ruleset, and all have H > 0*.

In each case, there is at least one incorrect cell state.  The failure to converge to the proper solution is indicated by the total energy -- all of these results have a total energy H = 1, whereas the ground state energy for proper solutions of the problem Hamiltonian have H = 0.  This conveniently spares us the effort of explicitly verifying the correctness of each cell.

Back-propagation of cellular automata using quantum annealing and adiabatic quantum computers may be of use in constructing initial conditions needed for universality proofs of additional CA rule sets, as well as other scenarios in which a desired output is known.

**Conway's Game of Life**

This approach is not restricted to 1-dimensional cellular automata.  A popular 2-D, 2-color cellular automata is Conway's Game of Life -- we now show that corresponding Hamiltonians can be constructed for AQC systems.

The generation rule of Conway's Game of Life is a nearest-neighbor ruleset, such that the evolution of a cell is determined by the number of neighbors in state = 1.

| neighbors with state = 1 | current state = 0 | current state = 1 |
|---|---|---|
| < 2 | 0 | 0 |
| 2 | 0 | 1 |
| 3 | 1 | 1 |
| > 3 | 0 | 0 |

*State transition rules for the Game of Life 2-D, 2-color cellular automata.*

Determining the count of a small number of qubits may be accomplished in several ways -- such as a variation of the Integer Weight Backpack problem [LUCAS] or the construction of a binary counter [DUNN]. For this problem, we choose the approach of Lucas with an additional qubit for the case of 0 living neighbors; although requiring more logical qubits, the QUBO coefficients for this particular problem have a smaller dynamic range than obtained using the binary counter and no additional logic is required to obtain comparison qubits tracking whether 2 or 3 neighbors are ‘alive’.

The Hamiltonian to count the number of living neighboring cells is:

$$H_{count} = \left(1 - \sum_{j=0}^{N} c_j\right)^2 + \left(\sum_i x_i - \sum_{j=0}^{N} j c_j\right)^2$$

where the ancillary qubits $c_j$ are Boolean variables such that $c_j = 1$ if the number of living neighboring cells is equal to j. The $x_i$ are the neighboring cell states.

We may now express the ruleset for Conway’s Game of Life as:

$$p_{n+1} = (p_n \wedge (c_2 \vee c_3)) \vee (\neg p_n \wedge c_3)$$

A Hamiltonian phrase to implement this rule for the evolution of cell $p_n$ to $p_{n+1}$ is given by the expression:

$$H_{term} = c_3 + 3p_{n+1} + p_n c_2 + 2p_n c_3 - 2p_n p_{n+1} + 3c_2 c_3 - 2c_2 p_{n+1} - 4c_3 p_{n+1}$$

The system Hamiltonian is constructed by summing the $H_{count}$ and $H_{term}$ expressions for all cells in all generations. As with the 1-D 2-color CA rules, this Hamiltonian is constructed in $O(N_{cells} * M_{generations})$ number of steps.

The choice of a two-dimensional CA ruleset does not directly impact the expression of the problem in qubit form -- the finite grid for a given generation can always be unwrapped into a 1D equivalent. However, 2-D CA systems have more nearest neighbors than do 1-D systems, and the increased inter-qubit connectedness makes embedding the Hamiltonian more problematic on low-connectivity hardware graphs. Interior cell-state qubits in middle generations in the Game of Life have interactions with 102 logical qubits (24 neighborhood cell qubits, 72 neighbors’ counting qubits, 2 of its own counting qubits in this generation, 2 of its own counting qubits in next generation, 1 ancestor, and 1 descendent), which greatly exceeds the connectivity of typical contemporary hardware qubit graphs. It is possible that a suitable gadget, incorporating both the CA ruleset and the counting logic, could be found that uses fewer ancillary qubits, but we did not pursue that search for this paper.

We adopt an edge-cell convention for the Game of Life study in which “off grid” cells are considered to be unpopulated. The 2-D nature of the cell pattern also drives a difference in the manner of representing each generation graphically -- we adopt the convention of sequential display of the cell grid, progressing from left to right.

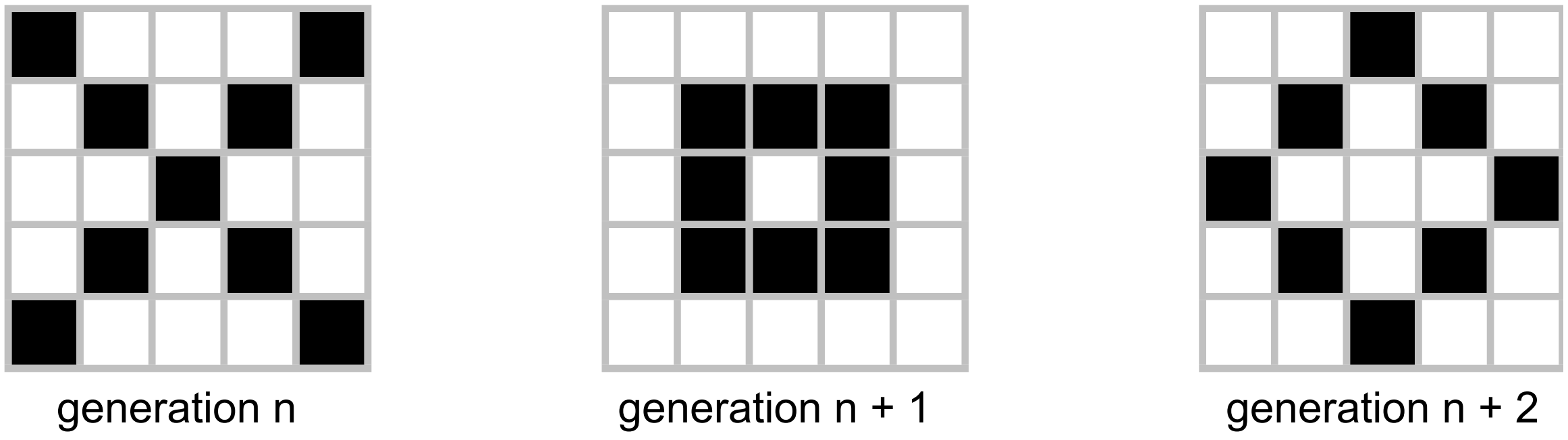

Figure 10: *An example of 3 successive generations of the Game of Life.*

The Game of Life has achieved popularity in recreational mathematics in part because it supports a number of mobile patterns and oscillating configurations. A search for oscillating patterns is particularly amenable to implementation in QUBO form, because a comparison of generations is easily constructed. Requiring successive generations *n* and *n+1* differ is equivalent to requiring that

$$\bigvee_{i \in cells} x_{i,n} \oplus x_{i,n+1} = 1$$

while requiring generation n and m to be identical is obtained via

$$\bigvee_{i \in cells} x_{i,n} \oplus x_{i,m} = 0$$

In this manner, oscillation searches of arbitrary length may be constructed. The generation-to-generation comparison terms are added to the CA ruleset Hamiltonian to obtain the overall system Hamiltonian.

Expressing the oscillating search terms in QUBO format is straight-forward, albeit requiring a number of ancillary qubits. An additional qubit is used to store the result of each cell-to-cell comparison. The native expression of the XOR assignment involves a 3-local term, so each cell-to-cell comparison requires an ancillary qubit to reduce the order to QUBO form, e.g.

$$C = A \oplus B \rightarrow H_{term} = A + B + C + 4D + 2AB - 2AC - 4AD - 2BC - 4BD + 4CD$$

Thus, for each generation in an N cell grid, 2N ancillary qubits are required for the cell-to-cell comparisons.

The generation-to-generation comparison is built from the cell-to-cell comparisons, which requires additional qubits for the expression of the OR logic. A typical approach is the pair-wise reduction of the constraint, thereby requiring O(N) additional logical qubits for an N cell grid -- although this may be reduced to O(log N) at the cost of a higher dynamic range of Hamiltonian coefficients.

An example of a 3-generation cycle is shown below:

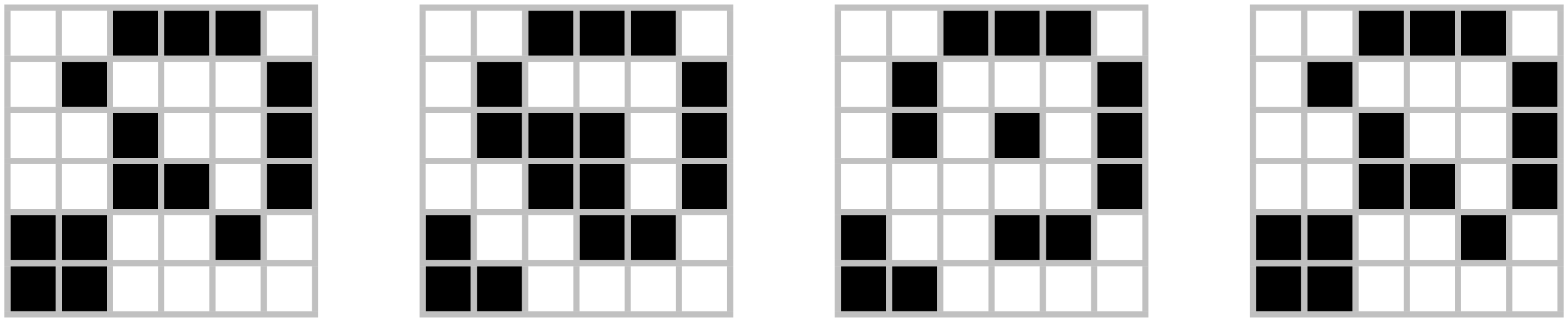

Figure 11*: A 3-step complete-graph oscillation cycle in the Game of Life, giving a repeating progression of A,B,C,A. (Identified via classical digital algorithm)*

Such oscillation searches are possible with digital computers, but the problem space scales exponentially with the size of the grid. In contrast, the number of logical qubits required to describe the problem as a QUBO grows linearly in the number of cells in the grid. If quantum annealing systems with hardware graphs capable of efficiently embedding QUBOs with this level of inter-qubit connectivity become available, it seems likely they would be an effective means of studying oscillating patterns in large grids.

**Hamiltonian Scaling**

Expressing these CA rules as QUBO models requires only a polynomial number of steps, needing O(N * M) steps for N cells over M generations. The number of logical qubits required is a function of the specific CA ruleset being studied, as some require more ancillary qubits than do others. However, the number of ancillary qubits required per cell is fixed, therefore the total number of logical qubits required scales in a predictable manner for any given rule.

The number of physical qubits required to embed the QUBO depends on both the rule set (which governs the inter-qubit connectivity of the QUBO) and the hardware graph of the quantum annealing system used. We are unaware of a general means of predicting the number of physical qubits required to embed an arbitrary QUBO, though the upper bound would certainly be given by the requirement for embedding a fully-connected qubit graph as given by [CHOI].

## The Need For Ancillary Qubits

Even in the case of 2-color, 1D, nearest-neighbor cellular automata, it is often the case that ancillary qubits are necessary to implement the desired behavior. Of the 256 elementary 1D, 2-color CA rulesets, we found 40 that can be implemented as QUBOs without using ancillary qubits (see Appendix A).

## Beyond Nearest-Neighbor Rules

There is nothing in the approach used herein that limits the study to that of nearest-neighbor cellular automata rule sets -- next-nearest-neighbor and larger rules may be implemented. However, because more cells are involved in each calculation, embedding the corresponding system Hamiltonian on a qubit hardware graph with limited connectivity may prove to be the limiting factor in studying those cases.

Similarly, there is nothing in this approach that constrains the choice of rules sets to only those described in 1-generation dependencies. Rules in which the CA evolution depends explicitly on prior generations (of a given cell, or of a cell's neighbors) can be constructed.

## Hybrid Quantum / Classical Solutions

The number of CA generations that may be simulated in a single run of an AQC system will depend on the number of physical qubits available. However, the nature of CA systems is such that the final generation state of one run may be used as the initial generation state of a subsequent run. It is therefore possible to have a driving program on a classical digital computer to orchestrate multiple AQC runs to perform a larger number of CA generations.

While that is doubtlessly an inefficient means of forward propagation of CA rules, such a hybrid solution may be of benefit when studying multi-generation backward propagations.

## Cellular Automata With More Than Two Colors

Quantum annealing systems and adiabatic quantum computers can be used to implement CA systems in which cells have more than two possible values. Doing so can be accomplished by the use of multiple logical qubits to represent a single cell -- specifically, using $\lceil \log_2 N \rceil$ logical qubits per cell for N colors.

Though the construction of the system Hamiltonian is more complicated, the same general approach can be used. We did not study systems with more than 2 colors as part of this paper.

**Further Research**

We have considered only CA rulesets in which the evolution of a cell's state can be expressed deterministically and in polynomial form as a function of that cell's state and history, along with the state and history of neighboring cells. Other, more complex, rule sets might be considered.

For convenience, all executions of the system Hamiltonians in this paper were performed on a digital computer. While successful at validating the behavior of the Hamiltonians, this approach does not address matters of hardware embedding or control precision which may impact the practicality of scaling to larger problems on current and near-term quantum annealing systems.

**Conclusions**:

Adiabatic quantum computers and quantum annealing systems can implement cellular automata rule sets on arbitrary input data sets (subject to available resources) provided the hardware supports at least 2-local terms in the system Hamiltonian. This includes rule sets known to be universal Turing machines, thereby confirming that AQC systems are themselves universal Turing machines. The Hamiltonians to implement the cellular automata behavior may be constructed in a polynomial number of steps for a given number of cells and generations in cases where the propagation rule can be distinctly expressed as a polynomial phrase for each cell.

AQC systems can determine rule-consistent unknown states in multiple generations, corresponding to an effective back-propagation of cellular automata.

While cellular automata are typically considered to be sequential in nature, we have shown that this is not an absolute constraint by the implementation of rule sets as multi-generational Hamiltonians. This raises the intriguing question as to which other nominally sequential algorithms may be implemented for adiabatic quantum computers by similar approaches.

**Appendix A**: Elementary 2-color, 1-dimensional, nearest-neighbor CA Rules that can be implemented as QUBO Hamiltonians without ancillary qubits. The rule numbering convention is that of [WOLFRAM].

Rule 0: $P_{n+1}$

Rule 3: $-L - P_n - P_{n+1} + L P_n + 2 L P_{n+1} + 2 P_n P_{n+1}$

Rule 5: $- L - R - P_{n+1} + L R + 2 L P_{n+1} + 2 R P_{n+1}$

Rule 10: $R + P_{n+1} - L R + 2 L P_{n+1} - 2 R P_{n+1}$

Rule 12: $P_n + P_{n+1} - L P_n + 2 L P_{n+1} - 2 P_n P_{n+1}$

Rule 15: $-L - P_{n+1} + 2 L P_{n+1}$

Rule 17: $- P_n - R - P_{n+1} + P_n R + 2 P_n P_{n+1} + 2 R P_{n+1}$

Rule 23: $- 2 L - 2 P_n - 2 R - 3 P_{n+1} + L P_n + L R + 2 L P_{n+1} + P_n R +$
$2 P_n P_{n+1} + 2 R P_{n+1}$

Rule 34: $R + P_{n+1} - P_n R + 2 P_n P_{n+1} - 2 R P_{n+1}$

Rule 43: $-L - P_n + 2 R - P_{n+1} + L P_n - L R + 2 L P_{n+1} - P_n R +$
$2 P_n P_{n+1} - 2 R P_{n+1}$

Rule 48: $L + P_{n+1} - L P_n - 2 L P_{n+1} + 2 P_n P_{n+1}$

Rule 51: $- P_n - P_{n+1} + 2 P_n P_{n+1}$

Rule 63: $- 2 L - 2 P_n - 3 P_{n+1} + L P_n + 2 L P_{n+1} + 2 P_n P_{n+1}$

Rule 68: $P_n + P_{n+1} - P_n R - 2 P_n P_{n+1} + 2 R P_{n+1}$

Rule 77: $- L + 2 P_n - R - P_{n+1} - L Pn + L R + 2 L P_{n+1} - P_n R$
$- 2 P_n P_{n+1} + 2 R P_{n+1}$

Rule 80: $L + P_{n+1} - L R - 2 L P_{n+1} + 2 R P_{n+1}$

Rule 85: $-R - P_{n+1} + 2 R P_{n+1}$

Rule 95: $- 2 L - 2 R - 3 P_{n+1} + L R + 2 L P_{n+1} + 2 R P_{n+1}$

Rule 113: $2 L - P_n - R - P_{n+1} - L P_n - L R - 2 L P_{n+1} + P_n R + 2 P_n P_{n+1} + 2 R P_{n+1}$

Rule 119: $- 2 P_n - 2 R - 3 P_{n+1} + P_n R + 2 P_n P_{n+1} + 2 R P_{n+1}$

Rule 136: $3 P_{n+1} + P_n R - 2 P_n P_{n+1} - 2 R P_{n+1}$

Rule 142: $P_n + R + P_{n+1} - L P_n - L R + 2 L P_{n+1} + P_n R - 2 P_n P_{n+1} - 2 R P_{n+1}$

Rule 160: $3 P_{n+1} + L R - 2 L P_{n+1} - 2 R P_{n+1}$

Rule 170: $R + P_{n+1} - 2 R P_{n+1}$

Rule 175: $- L + 2 R - P_{n+1} - L R + 2 L P_{n+1} - 2 R P_{n+1}$

Rule 178: $L + R + P_{n+1} - L P_n + L R - 2 L P_{n+1} - P_n R + 2 P_n P_{n+1} - 2 R P_{n+1}$

Rule 187: $- P_n + 2 R - 1 P_{n+1} - 1 P_n R + 2 P_n P_{n+1} - 2 R P_{n+1}$

Rule 192: $3 P_{n+1} + L P_n - 2 L P_{n+1} - 2 P_n P_{n+1}$

Rule 204: $P_n + P_{n+1} - 2 P_n P_{n+1}$

Rule 207: $-L + 2 P_n - P_{n+1} - L P_n + 2 L P_{n+1} - 2 P_n P_{n+1}$

Rule 212: $L + P_n + P_{n+1} + L P_n - L R - 2 L P_{n+1} - P_n R - 2 P_n P_{n+1} + 2 R P_{n+1}$

Rule 221: $2 P_n - R - P_{n+1} - P_n R - 2 P_n P_{n+1} + 2 R P_{n+1}$

Rule 232: $3 P_{n+1} + L P_n + L R - 2 L P_{n+1} + P_n R - 2 P_n P_{n+1} - 2 R P_{n+1}$

Rule 238: $P_n + R + P_{n+1} + P_n R - 2 P_n P_{n+1} - 2 R P_{n+1}$

Rule 240: $L + P_{n+1} - 2 L P_{n+1}$

Rule 243: $2 L - P_n - P_{n+1} - L P_n - 2 L P_{n+1} + 2 P_n P_{n+1}$

Rule 245: $2 L - R - P_{n+1} - L R - 2 L P_{n+1} + 2 R P_{n+1}$

Rule 250: $L + R + P_{n+1} + L R - 2 L P_{n+1} - 2 R P_{n+1}$

Rule 252: $L + P_n + P_{n+1} + L P_n - 2 L P_{n+1} - 2 P_n P_{n+1}$

Rule 255: $-P_{n+1}$

**Appendix B**: Hamiltonian table for Rule 110, as implemented herein.

The correct rule behavior is highlighted in ***italics***, which (as desired) has the global minimum Hamiltonian value. Results sorted by total energy *H*.

| L | $P_n$ | R | D | $C_1$ | $C_2$ | $P_{n+1}$ | $H_{term}$ |
|---|---|---|---|---|---|---|---|
| *0* | *0* | *0* | *0* | *0* | *0* | *0* | *0* |
| *0* | *0* | *1* | *0* | *0* | *1* | *1* | *0* |
| *0* | *1* | *0* | *0* | *0* | *1* | *1* | *0* |
| *0* | *1* | *1* | *0* | *1* | *1* | *1* | *0* |
| *1* | *0* | *0* | *0* | *0* | *0* | *0* | *0* |
| *1* | *0* | *1* | *0* | *0* | *1* | *1* | *0* |
| *1* | *1* | *0* | *0* | *0* | *1* | *1* | *0* |
| *1* | *1* | *1* | *1* | *1* | *1* | *0* | *0* |
| 0 | 0 | 0 | 0 | 0 | 0 | 1 | 1 |
| 0 | 0 | 0 | 0 | 0 | 1 | 1 | 1 |
| 0 | 0 | 1 | 0 | 0 | 0 | 0 | 1 |
| 0 | 0 | 1 | 0 | 0 | 1 | 0 | 1 |
| 0 | 0 | 1 | 0 | 1 | 1 | 1 | 1 |
| 0 | 1 | 0 | 0 | 0 | 0 | 0 | 1 |
| 0 | 1 | 0 | 0 | 0 | 1 | 0 | 1 |
| 0 | 1 | 0 | 0 | 1 | 1 | 1 | 1 |
| 0 | 1 | 1 | 0 | 0 | 1 | 1 | 1 |
| 0 | 1 | 1 | 0 | 1 | 1 | 0 | 1 |
| 0 | 1 | 1 | 1 | 1 | 1 | 0 | 1 |
| 1 | 0 | 0 | 0 | 0 | 0 | 1 | 1 |
| 1 | 0 | 0 | 0 | 0 | 1 | 1 | 1 |
| 1 | 0 | 0 | 1 | 0 | 0 | 0 | 1 |
| 1 | 0 | 1 | 0 | 0 | 0 | 0 | 1 |
| 1 | 0 | 1 | 0 | 0 | 1 | 0 | 1 |
| 1 | 0 | 1 | 1 | 0 | 1 | 0 | 1 |
| 1 | 0 | 1 | 1 | 1 | 1 | 0 | 1 |
| 1 | 1 | 0 | 0 | 0 | 0 | 0 | 1 |
| 1 | 1 | 0 | 0 | 0 | 1 | 0 | 1 |
| 1 | 1 | 0 | 1 | 0 | 1 | 0 | 1 |
| 1 | 1 | 0 | 1 | 1 | 1 | 0 | 1 |
| 1 | 1 | 1 | 0 | 0 | 1 | 1 | 1 |
| 1 | 1 | 1 | 0 | 1 | 1 | 1 | 1 |
| 1 | 1 | 1 | 1 | 1 | 1 | 1 | 1 |
| 0 | 0 | 0 | 0 | 0 | 1 | 0 | 2 |
| 0 | 0 | 1 | 0 | 0 | 0 | 1 | 2 |
| 0 | 0 | 1 | 0 | 1 | 0 | 0 | 2 |
| 0 | 0 | 1 | 0 | 1 | 1 | 0 | 2 |
| 0 | 0 | 1 | 1 | 1 | 1 | 0 | 2 |

| L | $P_n$ | R | D | $C_1$ | $C_2$ | $P_{n+1}$ | $H_{term}$ |
|---|---|---|---|---|---|---|---|
| 0 | 1 | 0 | 0 | 0 | 0 | 1 | 2 |
| 0 | 1 | 0 | 0 | 1 | 0 | 0 | 2 |
| 0 | 1 | 0 | 0 | 1 | 1 | 0 | 2 |
| 0 | 1 | 0 | 1 | 1 | 1 | 0 | 2 |
| 0 | 1 | 1 | 0 | 0 | 1 | 0 | 2 |
| 0 | 1 | 1 | 1 | 1 | 1 | 1 | 2 |
| 1 | 0 | 0 | 0 | 0 | 1 | 0 | 2 |
| 1 | 0 | 0 | 1 | 0 | 1 | 0 | 2 |
| 1 | 0 | 1 | 0 | 0 | 0 | 1 | 2 |
| 1 | 0 | 1 | 0 | 1 | 1 | 1 | 2 |
| 1 | 0 | 1 | 1 | 0 | 0 | 0 | 2 |
| 1 | 0 | 1 | 1 | 0 | 1 | 1 | 2 |
| 1 | 0 | 1 | 1 | 1 | 0 | 0 | 2 |
| 1 | 0 | 1 | 1 | 1 | 1 | 1 | 2 |
| 1 | 1 | 0 | 0 | 0 | 0 | 1 | 2 |
| 1 | 1 | 0 | 0 | 1 | 1 | 1 | 2 |
| 1 | 1 | 0 | 1 | 0 | 0 | 0 | 2 |
| 1 | 1 | 0 | 1 | 0 | 1 | 1 | 2 |
| 1 | 1 | 0 | 1 | 1 | 0 | 0 | 2 |
| 1 | 1 | 0 | 1 | 1 | 1 | 1 | 2 |
| 1 | 1 | 1 | 0 | 0 | 1 | 0 | 2 |
| 1 | 1 | 1 | 0 | 1 | 1 | 0 | 2 |
| 1 | 1 | 1 | 1 | 0 | 1 | 0 | 2 |
| 0 | 0 | 0 | 0 | 1 | 0 | 0 | 3 |
| 0 | 0 | 0 | 1 | 0 | 0 | 0 | 3 |
| 0 | 0 | 1 | 0 | 1 | 0 | 1 | 3 |
| 0 | 0 | 1 | 1 | 0 | 1 | 0 | 3 |
| 0 | 0 | 1 | 1 | 1 | 0 | 0 | 3 |
| 0 | 0 | 1 | 1 | 1 | 1 | 1 | 3 |
| 0 | 1 | 0 | 0 | 1 | 0 | 1 | 3 |
| 0 | 1 | 0 | 1 | 0 | 1 | 0 | 3 |
| 0 | 1 | 0 | 1 | 1 | 0 | 0 | 3 |
| 0 | 1 | 0 | 1 | 1 | 1 | 1 | 3 |
| 0 | 1 | 1 | 0 | 1 | 0 | 0 | 3 |
| 1 | 0 | 0 | 1 | 0 | 1 | 1 | 3 |
| 1 | 0 | 0 | 1 | 1 | 0 | 0 | 3 |
| 1 | 0 | 1 | 0 | 1 | 0 | 0 | 3 |
| 1 | 0 | 1 | 0 | 1 | 1 | 0 | 3 |
| 1 | 1 | 0 | 0 | 1 | 0 | 0 | 3 |
| 1 | 1 | 0 | 0 | 1 | 1 | 0 | 3 |
| 1 | 1 | 1 | 1 | 0 | 1 | 1 | 3 |
| 1 | 1 | 1 | 1 | 1 | 0 | 0 | 3 |
| 0 | 0 | 0 | 0 | 1 | 0 | 1 | 4 |

| L | $P_n$ | R | D | $C_1$ | $C_2$ | $P_{n+1}$ | $H_{term}$ |
|---|---|---|---|---|---|---|---|
| 0 | 0 | 0 | 0 | 1 | 1 | 1 | 4 |
| 0 | 0 | 0 | 1 | 0 | 1 | 0 | 4 |
| 0 | 0 | 0 | 1 | 1 | 0 | 0 | 4 |
| 0 | 0 | 1 | 1 | 0 | 0 | 0 | 4 |
| 0 | 0 | 1 | 1 | 0 | 1 | 1 | 4 |
| 0 | 1 | 0 | 1 | 0 | 0 | 0 | 4 |
| 0 | 1 | 0 | 1 | 0 | 1 | 1 | 4 |
| 0 | 1 | 1 | 0 | 0 | 0 | 0 | 4 |
| 0 | 1 | 1 | 0 | 1 | 0 | 1 | 4 |
| 0 | 1 | 1 | 1 | 0 | 1 | 0 | 4 |
| 0 | 1 | 1 | 1 | 1 | 0 | 0 | 4 |
| 1 | 0 | 0 | 0 | 1 | 0 | 0 | 4 |
| 1 | 0 | 0 | 1 | 0 | 0 | 1 | 4 |
| 1 | 0 | 0 | 1 | 1 | 1 | 0 | 4 |
| 1 | 0 | 1 | 0 | 1 | 0 | 1 | 4 |
| 1 | 1 | 0 | 0 | 1 | 0 | 1 | 4 |
| 1 | 1 | 1 | 0 | 0 | 0 | 0 | 4 |
| 1 | 1 | 1 | 0 | 1 | 0 | 0 | 4 |
| 0 | 0 | 0 | 0 | 1 | 1 | 0 | 5 |
| 0 | 0 | 0 | 1 | 0 | 1 | 1 | 5 |
| 0 | 0 | 0 | 1 | 1 | 1 | 0 | 5 |
| 0 | 1 | 1 | 0 | 0 | 0 | 1 | 5 |
| 0 | 1 | 1 | 1 | 0 | 1 | 1 | 5 |
| 1 | 0 | 0 | 0 | 1 | 0 | 1 | 5 |
| 1 | 0 | 0 | 0 | 1 | 1 | 1 | 5 |
| 1 | 0 | 0 | 1 | 1 | 1 | 1 | 5 |
| 1 | 0 | 1 | 1 | 0 | 0 | 1 | 5 |
| 1 | 0 | 1 | 1 | 1 | 0 | 1 | 5 |
| 1 | 1 | 0 | 1 | 0 | 0 | 1 | 5 |
| 1 | 1 | 0 | 1 | 1 | 0 | 1 | 5 |
| 1 | 1 | 1 | 0 | 0 | 0 | 1 | 5 |
| 1 | 1 | 1 | 0 | 1 | 0 | 1 | 5 |
| 1 | 1 | 1 | 1 | 0 | 0 | 0 | 5 |
| 0 | 0 | 0 | 1 | 0 | 0 | 1 | 6 |
| 0 | 0 | 0 | 1 | 1 | 1 | 1 | 6 |
| 0 | 0 | 1 | 1 | 1 | 0 | 1 | 6 |
| 0 | 1 | 0 | 1 | 1 | 0 | 1 | 6 |
| 1 | 0 | 0 | 0 | 1 | 1 | 0 | 6 |
| 1 | 0 | 0 | 1 | 1 | 0 | 1 | 6 |
| 1 | 1 | 1 | 1 | 1 | 0 | 1 | 6 |
| 0 | 0 | 0 | 1 | 1 | 0 | 1 | 7 |
| 0 | 0 | 1 | 1 | 0 | 0 | 1 | 7 |
| 0 | 1 | 0 | 1 | 0 | 0 | 1 | 7 |

| L | $P_n$ | R | D | $C_1$ | $C_2$ | $P_{n+1}$ | $H_{term}$ |
|---|---|---|---|---|---|---|---|
| 0 | 1 | 1 | 1 | 0 | 0 | 0 | 7 |
| 0 | 1 | 1 | 1 | 1 | 0 | 1 | 7 |
| 1 | 1 | 1 | 1 | 0 | 0 | 1 | 8 |
| 0 | 1 | 1 | 1 | 0 | 0 | 1 | 10 |